\documentclass[12pt]{article}
\usepackage{graphics,epsfig}
\usepackage[english]{babel}
\newcommand{\cinst}[2]{$^{\mathrm{#1}}$~#2\par}
\newcommand{\crefi}[1]{$^{\mathrm{#1}}$}

\begin{document}
%\linenumbers

%\begin{titlepage}

\thispagestyle{empty}
%%%%%%%%%%%%%%%%%%%%%%%% COVER PAGE
\begingroup

%\raisebox{0.5cm}[0cm][0cm] {
%\begin{tabular*}{\hsize}{@{\hspace*{5mm}}ll@{\extracolsep{\fill}}r@{}}
%\begin{minipage}[t]{3cm}
%\vglue.5cm
%\end{minipage}
%&
%\begin{minipage}[t]{7cm}

%\end{minipage}
%&
%\begin{minipage}[t]{7cm}
\vglue.5cm \hspace{9cm}{YerPhI Preprint - 1627 (2013)
%\end{minipage}
%\end{tabular*}
%}
\vglue.5cm
\begin{center}
%\hspace{9cm}{Draft 1}\\
%\hspace{9cm}{August 2013} \vglue 1.0cm

{\normalsize \bf  THE YIELDS OF DIFFERENT STATES 
 OF $\Delta(1232)$ \\[.3cm]IN NEUTRINO-INDUCED REACTIONS AT $\langle E_\nu \rangle \approx$ 10 GeV} 

\end{center}

\vspace{1.cm}
\begin{center}

 N.M.~Agababyan\crefi{1},
 N.~Grigoryan\crefi{2}, H.~Gulkanyan\crefi{2},\\
 A.A.~Ivanilov\crefi{3}, V.A.~Korotkov\crefi{3}

\setlength{\parskip}{0mm} \small
%\HRule\\

\vspace{1.cm} \cinst{1}{Joint Institute for Nuclear Research,
Dubna, Russia} \cinst{2}{Alikhanyan National Scientific Laboratory \\
(Yerevan Physics Institute), Armenia}
\cinst{3}{Institute for High Energy Physics, Protvino, Russia}
\end{center}

\setlength{\parskip}{0mm}
\small
%\HRule\\

\vspace{130mm}

{\centerline{\bf YEREVAN  2013}}

%\end{titlepage}
\newpage
\vspace{1.cm}

\begin{abstract}
First experimental data on the inclusive yields of $\Delta^{0}(1232)$ and  $\Delta^{+}(1232)$ isobars in neutrino-induced reactions are obtained. Their total yields in neutrino-nucleon interactions, 0.145$\pm$0.055 and $0.182\pm0.054$, respectively, exceed the value of $0.091\pm0.015$ obtained recently for the $\Delta^{++}(1232)$ state. The data for the all three states of $\Delta(1232)$ cannot be reproduced by the LEPTO6.5 model. The fractions of $\pi^0$, $\pi^-$ and $\pi^+$ mesons originating from the decay of lightest baryonic and mesonic resonances are estimated to be, respectively $56.6\pm9.6$\%, $39.5\pm6.6$\% and $27.7\pm3.2$\%.  
\end{abstract}

\newpage
\setcounter{page}{1}
\begin{center}
{\large 1. ~INTRODUCTION}\\
\end{center}

The space-time pattern of the leptoproduced quark string fragmentation incorporates the formation of both directly produced stable hadrons (pions, kaons, nucleons and antinucleons) and resonances (mesonic, baryonic and antibaryonic). The latter originates a significant fraction of final stable hadrons, as well as, to a much smaller extent, indirectly produced resonances. To infer as complete as possible information about the pattern of the quark string fragmentation, one needs as detailed as possible experimental data, not only on the leptoproduction of stable hadrons but also of hadronic resonances. At present, the experimental data are available for the total yields (for some cases, also for inclusive spectra) of light mesonic resonances up to the $f_2(1270)$ meson (see \cite{ref1,ref2} and references therein for the case of neutrinoproduction) and strange baryonic resonances $\Sigma^{*\pm}(1385)$ \cite{ref3, ref4,ref5,ref6,ref7,ref8}. The data on the non-strange baryonic resonances are  rather scarce and concern only $\Delta^{++}(1232)$  \cite{ref9, ref10,ref11}.

The aim of this work is to obtain the first experimental data on the inclusive neutrinoproduction $\Delta^0(1232)$ and $\Delta^+(1232)$. In Section 2, the experimental procedure is described. Section 3 presents the experimental data on the yields of $\Delta^0(1232)$, $\Delta^+(1232)$ and $\Delta^{++}(1232)$. A comparison with the LEPTO6.5 model prediction is performed. Section 4 is devoted to the estimation of the fraction of pions originating from the decay of lightest baryonic and mesonic resonances. The results are summarized in Section 5. 

\begin{center}
{\large 2. ~EXPERIMENTAL PROCEDURE}\\
\end{center}

The experiment was performed with SKAT bubble chamber \cite{ref12},
exposed to a wideband neutrino beam obtained with a 70 GeV primary
protons from the Serpukhov accelerator. The chamber was filled
with a propane-freon mixture containing 87 vol\% propane
($C_3H_8$) and 13 vol\% freon ($CF_3Br$) with the percentage of
nuclei H:C:F:Br = 67.9:26.8:4.0:1.3 \%. A 20 kG uniform magnetic
field was provided within the operating chamber volume.

Charged-current interactions containing a negative muon with
momentum $p_{\mu} >$0.5 GeV/c were selected. The overwhelming part of
protons with momentum below 0.6 GeV$/c$ and a fraction of protons  with momentum
0.6-0.85 GeV$/c$ were identified by their stopping in the chamber.
Stopping $\pi^+$ mesons were identified by their
$\pi^+$-$\mu^+$-$e^+$ decay. A fraction of low-momentum
($p_{\pi^+} < 0.5$ GeV$/c$) $\pi^+$ mesons were identified by the
mass-dependent fit provided that the $\chi^2$- value for the pion
hypothesis was significantly smaller as compared to that for
proton. Non-identified positively charged hadrons are assigned the pion mass or,
in the cases explained below, the proton mass.
It was required the errors in measuring the momenta be
less than 24\% for muon, 60\% for other charged particles and
$V^0$'s (corresponding to neutral strange particles) and less than 100\%
for photons. The mean relative error ($\Delta p/p$) in the momentum measurement for
muons, pions, protons and gammas was, respectively, 3\%, 6.5\%, 10\% and 19\%.
Each event was given a weight to correct for the
fraction of events excluded due to improperly reconstruction. More
details concerning the experimental procedure, in particular, the
estimation of the neutrino energy $E_\nu$ and the reconstruction of 
$\pi^0\rightarrow 2\gamma$ decays, can be found in our
previous publications \cite{ref13,ref14,ref15}.

Similarly to the case of light mesonic resonances \cite{ref2}, a restriction of $W > 1.8$ GeV 
was put on the invariant mass of the hadronic system, while no restriction was imposed on the transfer momentum squared $Q^2$. The number of accepted events was 
5242 (6557 weighted events). The mean values of kinematical variables ware $\langle E_\nu \rangle$ = 9.8 GeV, $\langle W \rangle$ = 2.8 GeV,  $\langle W^2 \rangle$ = 8.7 GeV$^2$, $\langle Q^2 \rangle$ = 2.6 (GeV/$c)^2$. 

The bubble-chamber methodics of this experiment enables, as it was shown in details in \cite{ref14,ref16}, to subdivide the whole sample of the selected events into two subsamples: the quasinucleon subsample $B_N$, corresponding to the interaction of neutrino with peripheral nucleons, and nuclear subsample $B_A$, corresponding to the
interaction of neutrino with arbitrary nucleons. A separate consideration of these two subsamples allows one to look for nuclear effects in the neutrinoproducion processes, as, for example, for the case of $\Delta^{++}(1232)$ production \cite{ref11} (see also related papers \cite{ref14,ref17,ref18}). The effective atomic weight of the composite nuclear target is estimated \cite{ref14} to be approximately equal to $A_{eff} = 21\pm2$.

In our study of the $\pi p$ effective mass distributions (presented in the next section) we included not only identified protons, but also non-identified positively charged hadrons (as protons), provided that the proton hypothesis for them was not rejected by the momentum-range relation in the propane-freon mixture. 
The experimental procedure for the selection of combination-candidates to the $\pi^+ p$ system is described in detail in \cite{ref10,ref11}, where first data on the inclusive neutrinoproduction of $\Delta^{++}(1232)$ in $\nu p$, $\nu n$, $\nu N$ (neutrino-nucleon) and $\nu A$ (neutrino-nucleus) interactions were obtained. As it was shown \cite{ref11}, the data on $\nu A$ interactions did not reveal any indication on the $\Delta^{++}(1232)$ production in the backward hemisphere in the laboratory frame (i.e. at $\cos \vartheta_{lab} < 0$, where 
$ \vartheta_{lab}$ is the ejection angle for the $\pi^+ p$ system with respect to the neutrino beam in the laboratory frame). Hence, in order to improve the resonance signal-to-background ratio, we required $\cos \vartheta_{lab} > 0$ for all combinations $\pi^- p$, $\pi^0 p$, $\pi^+ p$. Lastly, for the $\pi^0 p$ system, a cut $E_{\gamma} >$ 30 MeV on the photon energy was imposed and $\gamma\gamma$ candidates to $\pi^0$ meson were chosen in the mass window 0.09 $< m_{\gamma\gamma} < 0.18$ GeV/$c^2$. The $\pi^0 p$ effective mass distributions (presented in the next section) were corrected for the losses of the reconstructed $\pi^0$ mesons and contamination from the background $\gamma\gamma$ combinations.  
 
\begin{center}
{\large 3. ~EXPERIMENTAL RESULTS}\\
\end{center}

The effective mass distributions for the systems $\pi^- p$, $\pi^0 p$ and $\pi^+ p$ produced in $\nu N$ and $\nu A$ interactions are plotted in Figures 1, 2 and 3. They were fitted (similarly to \cite{ref9,ref10,ref11}) by a five parameter function

\begin{equation}
F(m) = C_1 \cdot BW(m) + C_2 \cdot BG(m) \,  ,
\end{equation}
\noindent where $BW(m)$ is the relativistic Breit-Wigner function \cite{ref19}, smeared according to
the experimental resolution for the each $\pi p$ system considered,while the background distribution was parametrized as

\begin{equation}
BG(m) = q^\alpha \cdot \exp(-\beta m^\gamma) \, ,
\end{equation}
\noindent where $q$ is the pion momentum in the $\pi p$ rest frame, and $C_1, C_2$, $\alpha, \beta, \gamma$ are fitted parameters. This parametrization provides a satisfactorily description for all distributions, except for the $\pi^- p$ system produced in $\nu A$ interactions. For the latter, a satisfactorily description was reached using a more complicated form for the $BG(m)$ function: 

 \begin{equation}
BG(m) = m_1^\alpha \cdot \exp(-\beta m_1-\gamma m_1^2-\delta m_1^3) \, ,
\end{equation}
\noindent where $m_1= m - m_\pi - m_p$ ($m_\pi$ and $m_p$ being the pion and proton masses, respectively), and $\alpha, \beta, \gamma, \delta$ are fitted parameters.

The fit results are shown in Figure 1, 2 and 3 and Table 1. The yields of $\Delta^0(1232)$ and $\Delta^+(1232)$ in Table 1 are corrected for the unobservable modes $n \pi^0$ and $n \pi^+$, respectively. 

We note first of all, that our data on $\Delta^{++}(1232)$ do not practically differ from those inferred recently in \cite{ref13} where no restriction on $W$ had been put. The inclusive yields of $\Delta^0(1232)$ and $\Delta^+(1232)$, quoted in Table 1, seem to dominate over the $\Delta^{++}(1232)$ yield. For instance, the ratio $R_N$ of their averaged total yield in $\nu N$ interactions, ${\langle n(\Delta^0 + \Delta^+) \rangle}_N /2 = 0.163\pm0.039$, to that for $\Delta^{++}(1232)$, ${\langle n(\Delta^{++}) \rangle}_N  = 0.091\pm0.015$, significantly exceeds unity: $R_N = 1.8\pm0.5$. The corresponding ratio for $\nu A$ interaction is equal to $R_A = 1.5\pm0.5$.  Unfortunately, the low statistical accuracy of our data does not allow one to infer the yields of $\Delta^0(1232)$ and $\Delta^+(1232)$ separately for $\nu p$ and $\nu n$ interactions and to perform a comparison with the corresponding yields of $\Delta^{++}(1232)$ measured in \cite{ref10}.

As for the case of the recent study \cite{ref11}, no statistically provided difference is observed in the $\Delta(1232)$ yields in $\nu N$ and $\nu A$ interactions, indicating that the nuclear effects (including both secondary intranuclear absorption and production processes) do not noticeably change the net yield of $\Delta(1232)$.

\begin{table}[ht]
\caption{The yields of $\Delta^0(1232)$, $\Delta^+(1232)$ and  $\Delta^{++}(1232)$ at three different ranges of the Feynman $x_F$ variable in  
$\nu N$ and $\nu A$ interactions at $W >$ 1.8 GeV.}
\begin{center}
\begin{tabular}{|l|c|c|c|}
  \hline
% &\multicolumn{4}{c|}{4 $< W^2 <$ 25
%GeV$^2$}\\ 
%\multicolumn{5}{|c|}{} \\ $h^+(x_F > 0)$
&$\Delta^0(1232)$&$\Delta^+(1232)$&$\Delta^{++}(1232)$ \\ 
 \hline
$\nu N$(all $x_F$)& 0.145$\pm$0.055&0.182$\pm$0.04&0.091$\pm$0.015 \\
$\nu A$(all $x_F$)& 0.163$\pm$0.065&0.132$\pm$0.040&0.100$\pm$0.019 \\ \hline
$\nu N$$(x_F < 0$)& 0.088$\pm$0.044&0.171$\pm$0.042&0.062$\pm$0.012 \\ 
$\nu A$$(x_F < 0$)& 0.145$\pm$0.060&0.146$\pm$0.039&0.070$\pm$0.018 \\ \hline
$\nu N$$(x_F > 0$)& 0.060$\pm$0.033&-0.019$\pm$0.031&0.025$\pm$0.011 \\ 
$\nu A$$(x_F > 0$)& 0.023$\pm$0.023&-0.005$\pm$0.023&0.031$\pm$0.009 \\ \hline
\end{tabular}
\end{center}
\end{table}
We also compared our data with the LEPTO6.5 model \cite{ref20} predictions for $\nu N$ interactions taking into account the proton and neutron contents of the composite nuclear target (at default values of the model parameters except those related to the experimental restrictions on the kinematical variables). As it has been shown recently \cite{ref13}, the experimentally measured yield of $\Delta^{++}(1232)$ in $\nu n$ interactions was in a reasonable agreement with the model predictions, which, however, badly overestimate (by about threefold) the yield of $\Delta^{++}(1232)$ in $\nu p$ interactions. As a result, this overestimation is retained also for $\nu N$ interactions (averaged over protons and neutrons of the composite nuclear target), for all three ranges of $x_F$ (the last column of Table 2). On the contrary, the model underestimates the yields of $\Delta^{0}(1232)$ in both three ranges of $x_F$, as well as the yield of  $\Delta^{+}(1232)$ at $x_F < 0$, at the same time predicting, in contradiction with the experiment, a rather moderate suppression of its yield at $x_F > 0$. We failed to reach a satisfactory description of our data by a unique set of the model input parameters. It should be also noted, that significant discrepancies between the LEPTO6.5 predictions (at the default values of the model parameters) and the experimental data were also reported for the cases of the neutrinoproduction \cite{ref6} and muonoproduction \cite{ref21} of strange baryonic resonances.
\begin{table}[h]
\caption{The yields of $\Delta^0(1232)$, $\Delta^+(1232)$ and  $\Delta^{++}(1232)$ at three different ranges of $x_F$ in $\nu N$ interactions, compared to the LEPTO6.5 model predictions (obtained at $W >$ 2 GeV).}
\begin{center}
\begin{tabular}{|l|c|c|c|}
  \hline
% &\multicolumn{4}{c|}{4 $< W^2 <$ 25
%GeV$^2$}\\ 
%\multicolumn{5}{|c|}{} \\ $h^+(x_F > 0)$
&$\Delta^0(1232)$&$\Delta^+(1232)$&$\Delta^{++}(1232)$ \\ 
 \hline
All $x_F$ &&&  \\
Experiment& 0.145$\pm$0.055&0.182$\pm$0.054&0.091$\pm$0.015 \\
Model&0.065&0.137&0.210 \\ \hline
$x_F < 0$&&&        \\ 
Experiment&0.088$\pm$0.044&0.171$\pm$0.042&0.062$\pm$0.012 \\ 
Model& 0.039&0.078&0.118 \\ \hline
$x_F > 0$& &&   \\ 
Experiment&0.060$\pm$0.033&$-$&0.025$\pm$0.011 \\ 
Model&0.026&0.059&0.092 \\ \hline
\end{tabular}
\end{center}
\end{table}

\begin{center}
{\large 4. ~THE FRACTION OF PIONS ORIGINATING FROM THE DECAY OF RESONANCES}\\
\end{center}

The obtained data on total yields of different states of $\Delta(1232)$ in $\nu A$ interactions (presented in Table 1), along with those for meson resonances \cite{ref2}, were used to estimate the fraction of $\pi^0$, $\pi^-$ and $\pi^+$ originating from the decay of lightest baryonic and mesonic resonances (the latter including 11 resonances from $\eta$ to $\Phi$). As it is seen from Table 3, these fractions are rather noticeable, being around $1/3\div1/2$. The decay fraction for neutral pions significantly exceeds that for charged ones.
%\newpage
\begin{table}[ht]
\caption{The mean multiplicities of pions (1st row) and the fraction (in \%) of pions originating from decay of  
$\Delta^0(1232)$, $\Delta^+(1232)$ and  $\Delta^{++}(1232)$ and lightest mesonic resonances.}
\begin{center}
\begin{tabular}{|l|c|c|c|}
  \hline
% &\multicolumn{4}{c|}{4 $< W^2 <$ 25
%GeV$^2$}\\ 
%\multicolumn{5}{|c|}{} \\ $h^+(x_F > 0)$
Resonance&$\pi^0$&$\pi^-$&$\pi^+$     \\
&0.904$\pm$0.066&0.652$\pm$0.010& 1.55$\pm$0.06\\ \hline
$\Delta^0(1232)$& 12.2$\pm$5.0&8.3$\pm$3.2&$-$ \\
$\Delta^+(1232)$&9.7$\pm$3.0&$-$&2.8$\pm$0.9 \\
$\Delta^{++}(1232)$&$-$&$-$&6.5$\pm$0.8 \\ 
 \hline
Light meson& 34.7$\pm$7.6&31.3$\pm$5.8&18.4$\pm$3.0 \\
resonances &&&\\ \hline
Sum&56.6$\pm$9.6&39.5$\pm$6.6&27.7$\pm$3.2 \\ \hline
\end{tabular}
\end{center}
\end{table}

\begin{center}
{\large 5. ~SUMMARY}\\
\end{center}

First experimental data on the inclusive yields of $\Delta^{0}(1232)$ and  $\Delta^{+}(1232)$ isobars in neutrino-induced reactions are obtained, both for $\nu N$ and $\nu A$ interactions. Their averaged total yield 
${\langle n(\Delta^0 + \Delta^+) \rangle}_N /2$ is estimated to be $0.163\pm0.039$ for $\nu N$ interactions and $0.148\pm0.038$ for $\nu A$ interactions, being noticeable larger that the corresponding yields of $\Delta^{++}(1232)$ state, $0.091\pm0.015$ and $0.100\pm0.019$, respectively. The LEPTO6.5 model predictions for $\nu N$ interactions underestimate the total yield of $\Delta^{+}(1232)$ and, more significantly, that of $\Delta^{0}(1232)$. On the other hand, the model badly overestimates the total yield of $\Delta^{++}(1232)$, as well as its yields at $x_F < 0$ and $x_F > 0$. Contrary to the model predictions, no $\Delta^{+}(1232)$ production is observed at $x_F > 0$.

No statistically provided difference is found in the $\Delta(1232)$ yields in $\nu N$ and $\nu A$ interactions, indicating that the nuclear effects (including both secondary intranuclear absorption and production processes) do not noticeably change the net yield of $\Delta(1232)$.

The fractions of $\pi^0$, $\pi^-$ and $\pi^+$ mesons originating from the decay of lightest baryonic and mesonic resonances are estimated to be, respectively $56.6\pm9.6$\%, $39.5\pm6.6$\% and $27.7\pm3.2$\%.  

\begin{center}
{\large ACKNOWLEDGMENTS}\\
\end{center}

%The authors from YerPhI acknowledge the supporting grants of
%Calouste Gulbenkian Foundation and Swiss Fonds Kidagan. 
The activity of one of the authors (H.G.) is supported by Cooperation
Agreement between DESY and YerPhI signed on December 6, 2002.

%%%%%%%%%%%%%%%%%%%%%%%%%%%%%%%%%%%%%%%%%%%%%%%%%%%%%%%%%%
   %%% References
%%%%%%%%%%%%%%%%%%%%%%%%%%%%%%%%%%%%%%%%%%%%%%%%%%%%%%%%%%

\newpage
\begin{figure}[ht]
\resizebox{0.9\textwidth}{!}{\includegraphics*[bb =20 65 600
610]{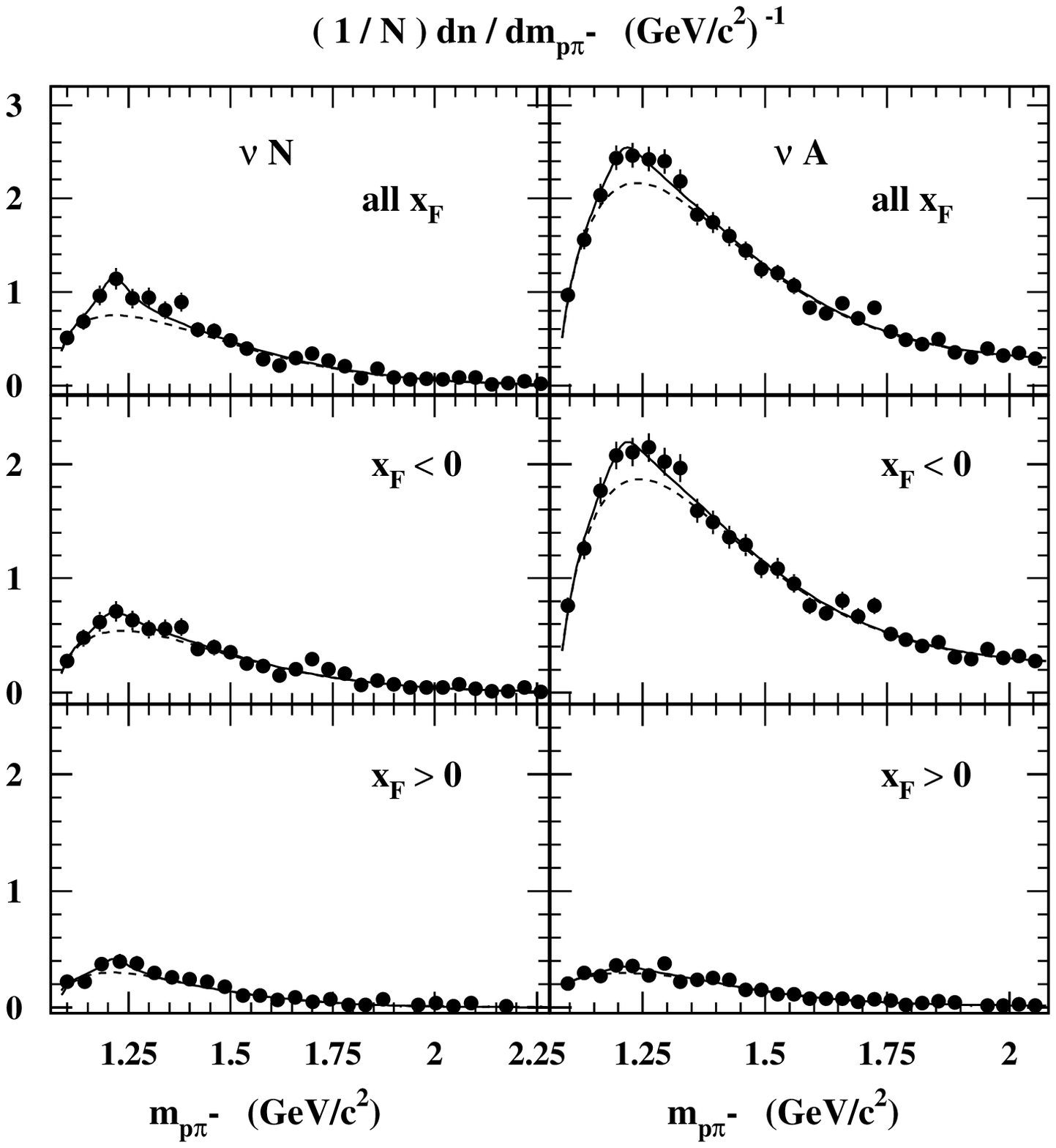}} \caption{The $\pi^- p$ effective mass
distributions for $\nu N$ and $\nu A$ interactions. The curves are
the fit result. The dashed curves correspond to the background distributions (see the text).}
\end{figure}

\newpage
\begin{figure}[ht]
\resizebox{0.9 \textwidth}{!}{\includegraphics*[bb=20 40 500 610]
{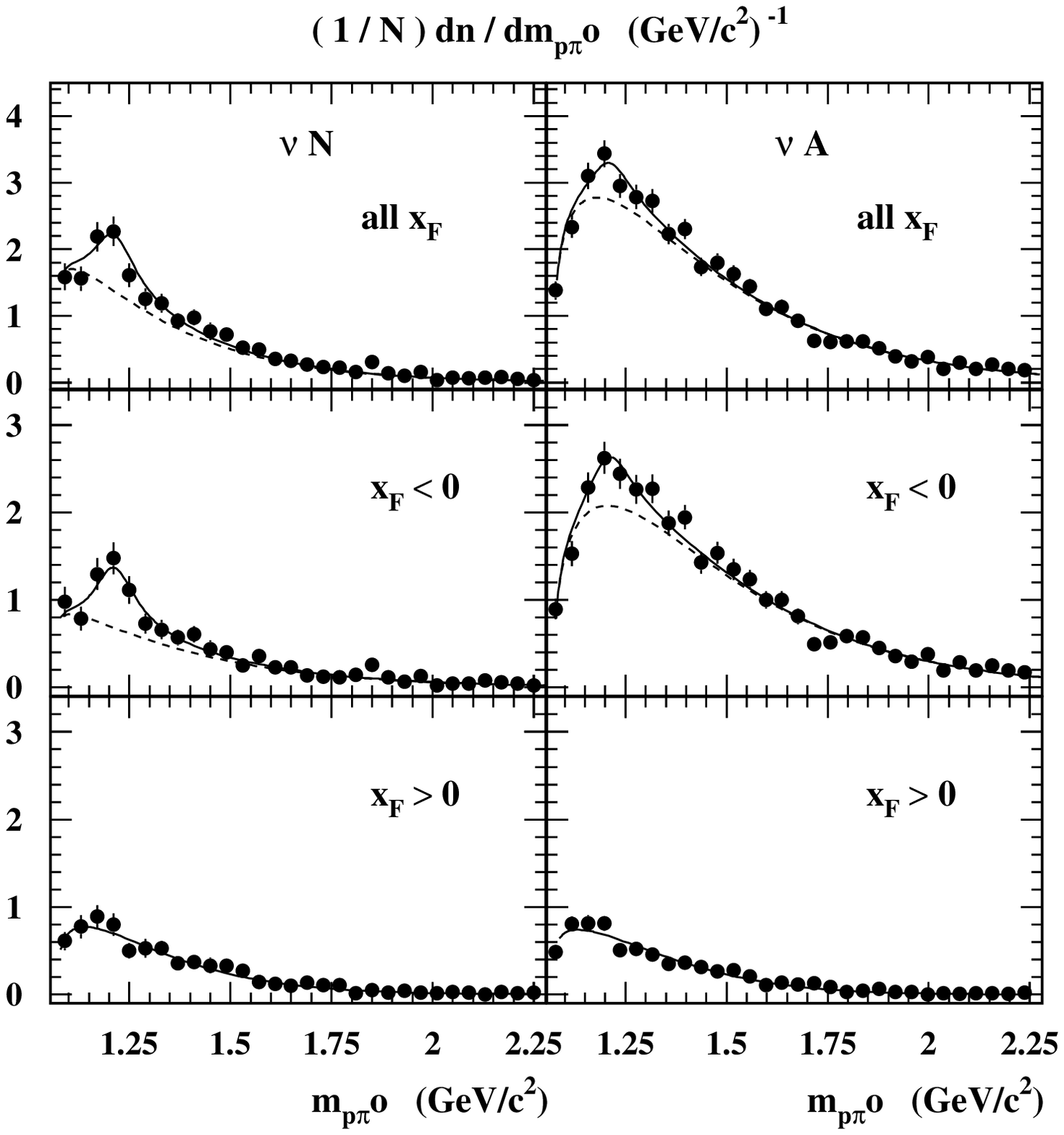}} \caption{The same as Figure 1, but for the $\pi^0 p$ system.}
\end{figure}

\newpage
\begin{figure}[ht]
\resizebox{0.9\textwidth}{!}{\includegraphics*[bb =20 65 600
610]{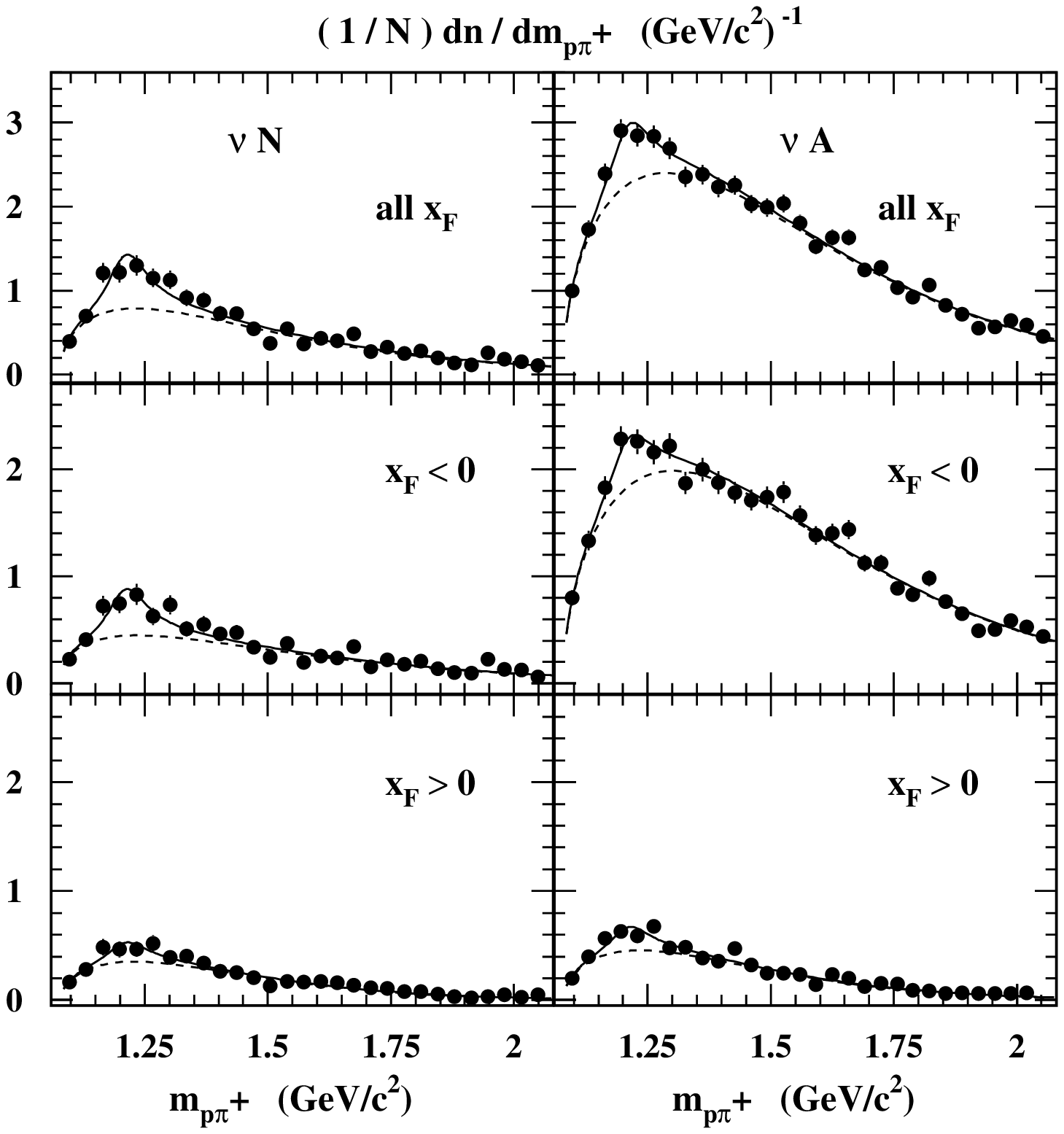}} \caption{The same as Figure 1, but for the $\pi^+ p$ system.}
\end{figure}

\end{document}